\documentstyle[epsfig]{aipproc}

\begin{document}

\title{
X-RAY OBSERVATION OF RADIO SUPERNOVA SN1979C BY ASCA 
}

\author{A. Ray$^{*)}$, R. Petre$^{\dagger)}$, E. Schlegel$^{\dagger\dagger)}$}

\address{$^{*)}$Tata Institute of Fundamental Research, Bombay 400 005, India;
E-mail: akr@tifr.res.in\\
$^{\dagger)}$Lab High Energy Astrophys., Goddard Space Flight Center,
Greenbelt, MD 20771\\
$^{\dagger\dagger)}$Smithsonian Astrophysical Observatory, 
60 Garden Street, Cambridge, MA 02138
}

%\authoraddr{LHEA Mail Code 661, NASA/GSFC, Greenbelt Road, MD 20771,
%U.S.A.}

\maketitle

\begin{abstract}
We report on the X-ray observation of the radio selected supernova
SN1979C carried out with ASCA in December 1997. The supernova of type
II$_{L}$ was first observed in the optical and occurred in the weakly
barred, almost face on spiral galaxy NGC 4321 (M100)
which is at a distance of 17.1 Mpc, and
contains at least three other supernovae discovered in this
century. 
%The Solid-State Imaging Spectrometer (SIS) had an exposure of
%27.3 ks while the Gas Scintillation Imaging Spectrometer had an
%exposure of 30.3 ks. 
No point source was detected at the radio
position of SN1979C in a 3' diameter half power response circle in a
27.3 ks SIS exposure. The
background and galaxy subtracted SN signal had a 3$\sigma$ upper limit
to the count rate of 1.2$\times 10^{-3}$ cps in the full ASCA SIS band
(0.4-10.0 keV).  These measurements give the first ever x-ray flux
limit of a Type II$_{L}$ SN in the higher energy band ($\geq$ 2 keV)
which is an important diagnostic of the {\it outgoing} shock wave
ploughing through the circumstellar medium.  
%The ROSAT
%HRI detected the SN in 1995 at an X-ray luminosity
%level of $L_x = 1.0 \times ^{39} \rm erg \; s^{-1}$ (in the 0.1-2.4
%keV band), while an observation by the Einstein HRI in 1979-1980 with
%a combined 41.3 ks exposure, did not detect the SN. 

\end{abstract}

%\keywords{stars:supernovae -- supernovae:individual (SN1979C)

%--  galaxies:individual (NGC4321) -- X-ray:stars} 

%\vsize=8.5 true in

\def \refhang{\noindent \hangafter=0 \hangindent=15pt }

%\centerline{}

\section{X-ray and Radio emission from Supernovae}

The observation of X-rays provides the most direct view of the region
of 
interaction between the ejecta from a core-collapse supernova (SN) and
the circumstellar medium of the progenitor star (see e.g. Chevalier \&
Fransson 1994).  This circumstellar interaction gives rise
to copious X-ray emission within days to months of the supernova
explosion, 
in contrast to X-ray emission that
occurs in the later supernova remnant (SNR) during the free expansion,
adiabatic or radiative phases.  The circumstellar matter is heated by
the
outgoing SN shock wave while the SN ejecta are heated by a reverse
shock wave.
At the same time, this interaction is expected to produce nonthermal
radio radiation; in recent years several young SNe have
been found to be emitting in the radio bands (see Weiler et
al. 1998). Relativistic electrons and magnetic fields in the region,
which give rise to the radio emission, may be built up by acceleration
in the shock wave and field amplification through Rayleigh-Taylor
instabilities. The detection of strong radio emission from a SN
provides good empirical evidence of circumstellar interaction of
ejecta, and such a SN is a probable emitter at X-ray wavelengths as
well.
%
%Until now there have been detections of nine young extragalactic SNe in
%the
%X-ray band, all of which are classified as type II SNe on the basis of
%optical spectra.  These are: SN1978K, SN1979C, SN1980K, SN1986J,
%SN1987A, SN1988Z, SN1993J, SN1994I and SN1995N.
%While the outgoing shock wave probes the circumstellar
%medium, the reverse shock wave probes the structure of the atmosphere
%of the exploding star, as it propagates into the ejecta.  
%X-ray emission from SNe that turns on shortly after the explosion and
%remains active for a long time is most likely due to the thermal
%emission from the hot gas produced in the interaction between the
%shock wave from the explosion and the circumstellar medium.  X-ray
%emission with a temperature of $\approx$ 1 keV (which falls in the
%ASCA spectral band) results from a reverse shock that propagates back
%into the expanding debris. This in turn is generated by the outgoing
%shock sweeping up circumstellar matter as it expands outwards --
%thereby creating an increasing ram pressure to drive the reverse shock
%(Chevalier \& Fransson, 1994).  
The temperature behind the outgoing
shock is typically 10 keV and therefore the bulk of this radiation
detected by ASCA would appear in its higher energy bands.  In the
circumstellar interaction model, the shocked supernova ejecta is the
dominant source of X-rays because of its higher density and lower
temperature.

\section{Observed properties of SN1979C by ASCA and previous missions}

%SN1979C was discovered in the optical by Johnson (1979) near
%maximum optical light (m$^{max}_B \leq 12$) on April 19, 1979 
%in the weakly-barred, almost face-on spiral galaxy NGC~4321
%(M100).  This galaxy is a member of the Virgo S cluster, at a distance
%of 17.1
%Mpc (Freedman et al 1994). Three other SNe have been observed in this
%galaxy this century: SN1901B, SN1914A, and SN1959E.  The J2000 position
%of the SN is at $\alpha$ = $12^h 22^m 58^s.58$, $\delta$ = $+ 15^{o}
%47^{'} 52^{s}.7$, offset to the southeast from the center of NGC~4321
%by $\sim$100 arc seconds.  
The supernova is thought to have originated from a 
red supergiant
progenitor with M$_{ZAMS} = 17-18 \pm 3
M_{\odot}$ (Van Dyk et al. 1999).  
An observation by the Einstein HRI in December 1979 established only
a 3$\sigma$ upper limit of $L_x \leq 8.8 \times 10^{39} \rm \; erg \;
s^{-1}$
(Palumbo et al.
1981).  An analysis of the combined Einstein data (of total duration
41.3 ks) between days 64 and 454 after explosion gave an upper limit
of 5.9 $\times 10^{39} \rm erg \; s^{-1}$ (Immler et al. 1998,
assuming
a 5 keV coronal plasma spectrum). The foreground column density
of neutral hydrogen is N$_H = 2.3 \times 10^{20} \rm cm^{-2}$ 
(Immler et al 1998).
An X-ray observation by the ROSAT HRI detected a source coincident
with the position of SN1979C for the first time, more than 16 years
after the explosion (source H25, Immler et al. 1998).  The corrected
ROSAT HRI position is at $\alpha = 12^h 22^m 58^s.57$, $\delta = +
15^{o} 47^{'} 53.5^{"}$ (J2000), within about 2.7$\arcsec$ with the
position determined by radio interferometry (Penhallow 1980).  The
measured count rate is $6.9 \times 10^{-4} \rm \; cps$. The
corresponding 0.2 - 2.4 keV flux and
luminosity are $F_x = 2.9 \times 10^{-14} \rm \; erg \; cm^{-2} \;
s^{-1}$ and $L_x = 1.0 \times 10^{39} \rm \; erg \; s^{-1}$
respectively, assuming a 5 keV thermal spectrum.

In this paper, we report on an X-ray observation of SN1979C using
ASCA.  These observations provide the first constraint on the X-ray
flux
in the energy band higher than 2 keV for a Type II$_{L}$ SN.
NGC 4321 was observed by ASCA on December 18-19, 1997.  The Solid State
Imaging Spectrometers (SIS) had useful exposure times of 24.3 ks (SIS0)
and 25.2 ks (SIS1); the Gas 
Scintillation Imaging Spectrometers (GIS) had exposure times of 27.5 ks
(GIS2) and 28.1 ks (GIS3).
%The nominal pointing position, $\alpha$ = $12^h 23^m 58^s.16$, $\delta$
%= $+ 15^{o} 51^{'}$, placed the galaxy and the supernova towards the
%center of the best calibrated SIS chips (chip 1 of SIS0 and chip 3 of
%SIS1).
%
A combined SIS0+SIS1 image is shown in Figure 1.
%while one for GIS2+GIS3 is shown in Figure 2.  
A 3$^{\arcmin}$
diameter half power response circle is shown superposed at the location
of SN1979C. Emission associated with the known agglomoration of sources
in the central region of NGC~4321 is clearly seen.  On the other hand,
no source is evident in the circle.  A point source detection task
failed to detect SN1979C in either the SIS or GIS image.

%
%\section{Results}
%
Placing an upper limit on the ASCA band flux from SN1979C is
complicated by contamination from the integrated flux from the sources
in the nuclear region of NGC 4321.  The center of NGC 4321 is only 100
arc seconds from SN1979C, resulting is a non-negligible overlap of
point responses.  Furthermore, the emission from the galaxy is clearly
broader than that produced by a single unresolved source.  Thus
determining a count rate from SN1979C requires subtraction both of
diffuse background and the contribution of the integrated flux from the
other sources in the galaxy.

%A 23926 s data set of SIS0 detector showed the background subtracted
%counts in the 3$^{\arcmin}$ half power circles around the galaxy
%and SN to be 178 $\pm$ 18 counts and 82 $\pm$ 15 counts respectively.
%Therefore the background subtracted counts in
%the said half power circles around the galaxy and SN are 178 $\pm 18$
%counts and 82 $\pm 15$ counts respectively.  As the estimated overlap
%between the two circles centered about the galaxy and SN is estimated
%to be $\sim$0.4, the background- and galaxy-subtracted SN signal is 11
%$\pm 16$ counts in the full 0.4 - 10 keV band.  The implied 3$\sigma$
%upper limit of the count rate is: 2.0 $\times 10^{-3}$ cps. A similar
An analysis of SIS1 data of 25 ks exposure, combined with the above 24 ks
exposure for SIS0, gives a combined upper limit to the count rate of
1.2 $\times 10^{-3}$ cps. The corresponding upper limit to the
combined data from GIS2 and GIS3 (28 ks each) gives 8 $\times 10^{-4}$
cps.
The net exposures, total galaxy- and background-subtracted net source
counts and their 1 $\sigma$ errors together with the 3$\sigma$ upper
limits on the SN count rates for SIS (S0 and S1) and GIS (G2 and G3)
are given in Table 1.
The above SIS combined upper limit corresponds to a flux limit of $3.5
\times 10^{-14} \; \rm erg \; cm^{-2} \; s^{-1} $ in the 0.4-10.0 keV
band for a Thermal Bremsstrahlung spectrum with T = 3 keV ($N_H = 3
\times 10^{20} \rm cm^{-2}$).
%
%
%We note that this reported luminosity limits are lower than that
%observed for SN1987A so far (Itoh et al. 1997).  The measured flux for
%SN1987A implies a 0.5-5 keV luminosity of $\sim$3$\times$10$^{39}$
%ergs s$^{-1}$ at the distance of SN1979C.  The corresponding ROSAT HRI
%flux from the measurement of Immler et al (1998) in the 0.5-2.0 keV
%band (for the same spectrum and absorption column for SN1979C) is $2
%\times
%10^{-14} \rm erg \; cm^{-2} \; s^{-1}$ which corresponds to a
%luminosity of $\sim$7$\times$10$^{38}$ ergs s$^{-1}$.  
%Note that for
%the energy bands in common between the two satellites ($\sim$0.5-2.0
%keV), ROSAT provides the more sensitive measure.
%
In Table 2 we give the detected x-ray source fluxes (or the upper
limits) for three separate missions Einstein, ROSAT and ASCA along
with their respective band-passes and luminosities at different
epochs.

\section{Discussion}

Until now almost all SNe detected in the X-ray were measured in the
lower energy band (0.1 - 2.0 keV) of ROSAT.  Our reported measurement
of SN1979C here gives the flux limit in the higher energy band ($>$ 2
keV) for the first time for a Type II$_{L}$ SN (the only other X-ray
detected Type II$_{L}$ SN is SN1980K).  Because the outgoing shock
wave
expanding through circumstellar material generally creates a higher
temperature emission spectrum compared to the reverse shock wave, the
measurement of SN X-ray fluxes in the higher energy bands constitutes
as important goal.

Both the ROSAT detection of the x-ray
emission as well as our ASCA upper limit referred to above are well
below its expected value if we scale its X-ray emission with respect to
another
type II-L SN active in the radio and X-ray, e.g. SN1980K. 
This lower than expected $L_x / L_R$ may
mean that the X-ray and radio emission properties of SNe could be
quite variable from one environment to another, i.e. the emission in
the two bands are not directly scalable in a universal way or that the
X-ray emission could be more strongly time variable than the radio
emission. The shock may be interacting with local cloudlets
to give enhanced thermal X-ray emission at certain epochs whereas the
radio radiation may be coming from a more global arrangement of
amplified magnetic fields and synchrotron radiating accelerated
electrons with long lifetimes.

%The detection probability of a SN against a bright galactic background
%may be limited by a given telescope's half power diameter (HPD),
%especially when a SN fades with time. The higher spatial resolution of
%the new generation of x-ray telescopes 
%will have improved capability in this regard and will also be
%important for their higher energy sensitivity required for both
%outwardly propagating and reverse shockwave thermodynamics.
%
%\section{Acknowledgements}
%
One of us (A.R.) was a Senior Research Associate of the National
Research 
Council at NASA/Goddard Space Flight Center during the course of this
work
and acknowledges support from the NRC and ASCA Guest Observer Program.
He also thanks the Aspen Center for Physics for hospitality where this
manuscript was completed.

\begin{table}
%\begin{center}
\caption{ASCA observation of SN 1979C in NGC4321 in December 1997}
\label{T:aray:1}
%\begin{tabular}{lrrr}
%\begin{tabular}{lccc}
\begin{tabular}{cccccc}
\hline
\hline
 & & & Net source \\
   Instrument& Exposure (ks) & Band (keV) &
   \multicolumn{1}{c}{counts} &
   \multicolumn{1}{c}{Counts/sec (3$\sigma$ lim)} & Flux\\
\tableline
\cr
SIS: S0 + S1 & 24 + 25 & 0.4-10.0 & $29.6 \pm 20$ & $\leq 1.2 \times
10^{-3}$ & $\leq$6.3$\times$10$^{-14}$ \\

             &         & 2.0-10.0 & $13.3 \pm 11$ & $\leq 6.7 \times
10^{-4}$ & $\leq$3.5$\times$10$^{-14}$\\

GIS: G2 + G3 & 28 + 28 & 1.0-10.0 & $-3.5 \pm 16$ & $\leq 8.4 \times
10^{-4}$ & $\leq$4.4$\times$10$^{-14}$ \\

             &         & 2.0-10.0 & $-12.0\pm 11$ & $\leq 6.2 \times
10^{-4}$& $\leq$3.2$\times$10$^{-14}$ \\
\tableline
\end{tabular}
%\end{center}
\end{table}

%\newpage

%%% 209

\begin{table}
\caption{SN 1979C X-ray flux and upper limits from different missions}
\label{T:aray:2}
%\begin{tabular}{lrrr}
%\begin{tabular}{lccc}
\begin{tabular}{ccccc}
\hline
\hline
 Mission   & Date & Exp (ks) &
   \multicolumn{1}{c}{Band (keV) 
} &
   \multicolumn{1}{c}{Flux ($\rm erg \; cm^{-2} \; s^{-1}$)}\\
\tableline
\cr

Einstein HRI & Dec 1980 & 41 &  0.1 - 4.5 & $\leq 1.7 \times 10^{-13}$
\\

ROSAT HRI & Jun 1995 & 42.8 &  0.1 - 2.4 & $2.0 \times 10^{-14}$ (TB
3keV) \\

ASCA  SIS & Dec 1997 & 49 & 2.0 - 10.0 & $\leq 3.5 \times 10^{-14}$ (TB
3keV) \\

%\tableline
\end{tabular}
\end{table}

%{\hbox{\vskip -0.5truein}}
\begin{figure} [b!] % fig 1
\centerline{\epsfig{file=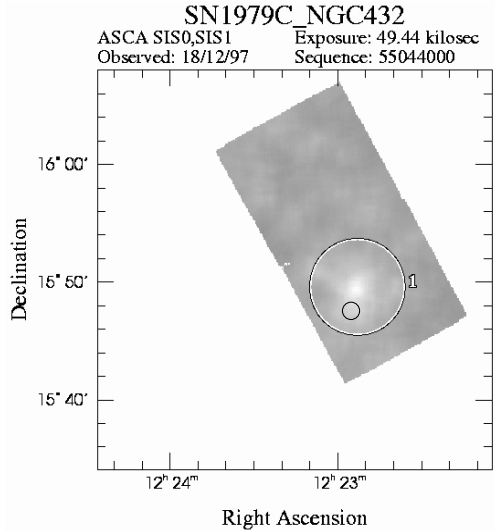,height=2.8in,width=2.6in}}
\vspace{2pt}
\caption{
ASCA SIS image of the galaxy NGC4321 (M100) containing the SN1979C
}
\label{F:aray:1}
\end{figure}

%\begin{figure}[b!] % fig 2
%\centerline{\epsfig{file=eric_sn79c_final_gis.eps,height=7.3in,width=4.6in}}
%\vspace{5pt}
%\caption{
%ASCA GIS image of the galaxy NGC4321 (M100) containing SN1979C
%}
%\label{F:aray:2}
%\end{figure}

\end{document}